\documentclass[]{tPHM2e}

\begin{document}

\markboth{Quilliet {\it et al.}: disorders in foams}{}

\title{\bf Topological and geometrical disorder correlate robustly in two-dimensional  foams}
\author{C. Quilliet $^{\rm a}$ $^{\ast}$
\thanks{$^\ast$
Corresponding author. 
Tel : +33 4 76 51 43 33. Fax : +33 4 76 63 54 95. Email: catherine.quilliet@ujf-grenoble.fr
\vspace{6pt}
} 
S. Ataei Talebi $^{\rm a}$, 
D. Rabaud $^{\rm a}$, 
J. K\"{a}fer  $^{\rm a}$, 
S. J. Cox  $^{\rm b}$, 
 and F. Graner  $^{\rm a}$, 
 \\
 \vspace{6pt}  
$^{\rm a}${\em{
Laboratoire de 
Spectrom\'{e}trie Physique$^{\ast\ast}$, B.P. 87, F-38402 Martin
d'H\`{e}res Cedex, France
\thanks{$^{\ast\ast}$
UMR5588, CNRS-Universit\'{e} Grenoble I
\vspace{6pt}
} 
}}; $^{\rm b}${\em{
Institute of Mathematics and Physics, Aberystwyth University, Ceredigion SY23 3BZ, UK
}}\\\vspace{6pt}}

\date{\today}

\maketitle

\begin{abstract}
A 2D foam can be characterised by its distribution of bubble areas, and of number of sides. Both distributions have an average and a width (standard deviation).  There are therefore at least two very different ways to characterise the disorder. The former is a geometrical measurement, while the latter is purely topological.  We discuss the common points and differences between both quantities. We measure them in a foam which is sheared, so that bubbles move past each other and the foam is ``shuffled" (a notion we discuss). Both quantities are strongly correlated; in this case (only) it thus becomes sufficient to use either one or the other
to characterize the foam disorder. We suggest applications to the analysis of other systems, including biological tissues.
\bigskip

\bigskip

\end{abstract}

\section{Introduction}

Two-dimensional cellular patterns fill the plane without gaps or overlaps. Even if we consider here only the class of patterns where most cells meet in threes, they are ubiquitous in nature:
from 2D foams to biological tissues, including geology, hydrodynamics, ecology, geography; as well as 2D cuts of 3D cellular patterns in metallurgy, astronomy, 3D foams and biological organisms  \cite{weairerivier,glazier}. Each cell is characterised by its area $A$, which  is an intrinsic property determined by the amount of matter it encloses, and its number of neighbours $n$, which can vary according to the mutual arrangement of cells within the pattern.

There are at least two quantities which characterise the disorder of a pattern: the distribution of $A$ and the distribution of $n$  \cite{weairerivier,glazier,glazierweaire}. 
In the present paper, we study whether it is possible to simplify the description of disorder to only two numbers, namely the width of each distributions, called, respectively, the geometrical  and topological disorders.  
Both measures of disorder are independent \cite{teixeira,granerjjf01}.
However, statistically, a cell's value of $n$ tends to increase with its size (perimeter or area):
small cells tend to have fewer neighbours, and larger cells have more neighbours
(see \cite{weairerivier,glazier,lewis,smith,lisso,pina,cargese,bideau,fortes} and references therein).
We can thus expect that, except when the initial conditions due to preparation are important, there is a correlation between these two measures of disorder. We investigate here, on different  examples but focusing on a 2D foam, whether this intuition is correct  and under which conditions the disorder of a pattern can statistically be described by a {\em single} number.

We first describe experiments on a sheared monolayer of bubbles (\S 2),
 and compare the results (\S 3) with simulations of similar systems and
 with data from evolving biological tissues (\S 4).

\section{Materials and methods}

\subsection{Foam preparation}

\begin{figure}
(a) \includegraphics[width=4cm]{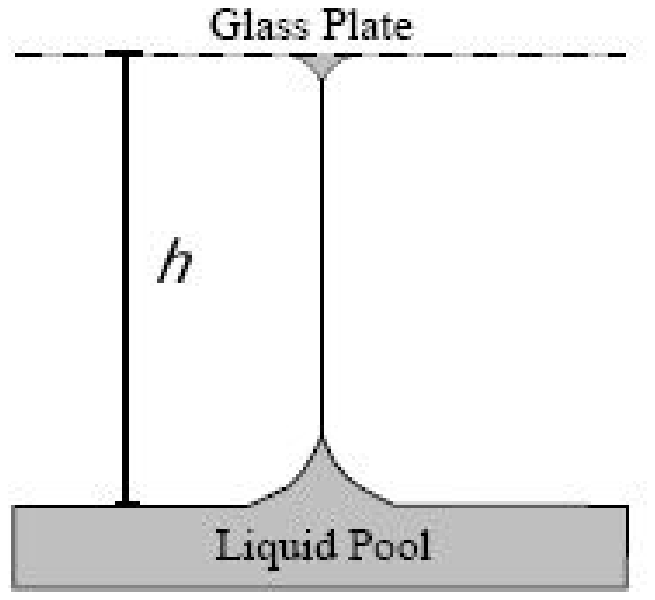}
(b) \includegraphics[width=7cm]{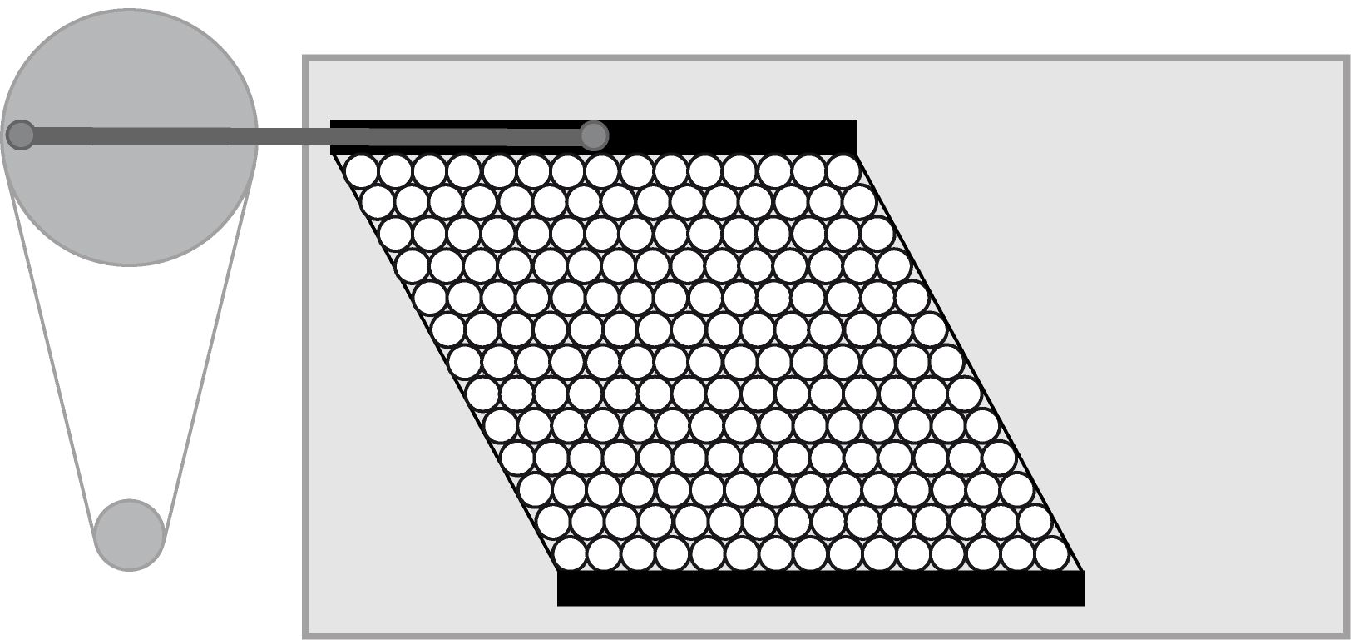}
\caption{Experimental set-up. (a) Side view: bubbles are sandwiched between the surface of water and a horizontal glass plate. (b) Top view: a computer controlled motor moves one rigid boundary, while the other one is fixed; the lateral boundaries are passive.}
\label{setup}
\end{figure}

The set-up is similar to that of  Kader and Earnshaw \cite{kader}.
It uses a bubble monolayer (quasi-2D foam) confined at the air-water interface (Fig. \ref{setup}a), as introduced by Smith \cite{smith} and adapted by Vaz \& Fortes \cite{vazfortes}. It provides all the properties that we require (unlike both other
 main types of quasi-2D foams, namely between two parallel glass plates, or at the surface of water exposed to air  \cite{vazcox}):
the water provides easy access to the foam and enables us to prepare precisely the chosen  distribution of bubble areas. It also facilitates the manipulation of the boundaries, in particular that of the rubber bands (see below). The glass plate facilitates observation and prevents bubbles from breaking: the foam lifetime is limited only by coarsening, which takes several hours and does not affect the results presented below.

\begin{figure}
{\small a)}
\includegraphics[width=4.7cm]{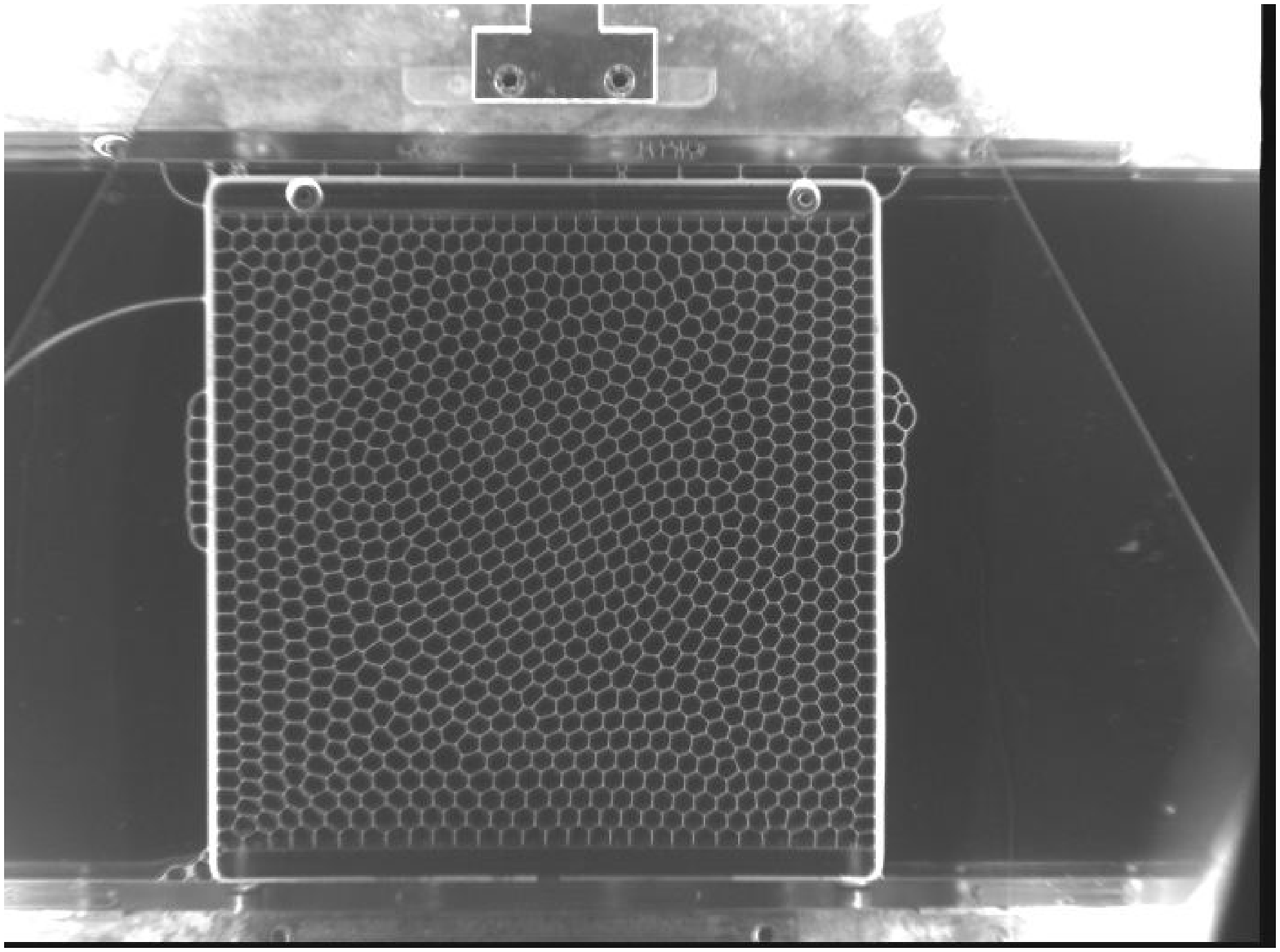}
{\small b)}
\includegraphics[width=4.7cm]{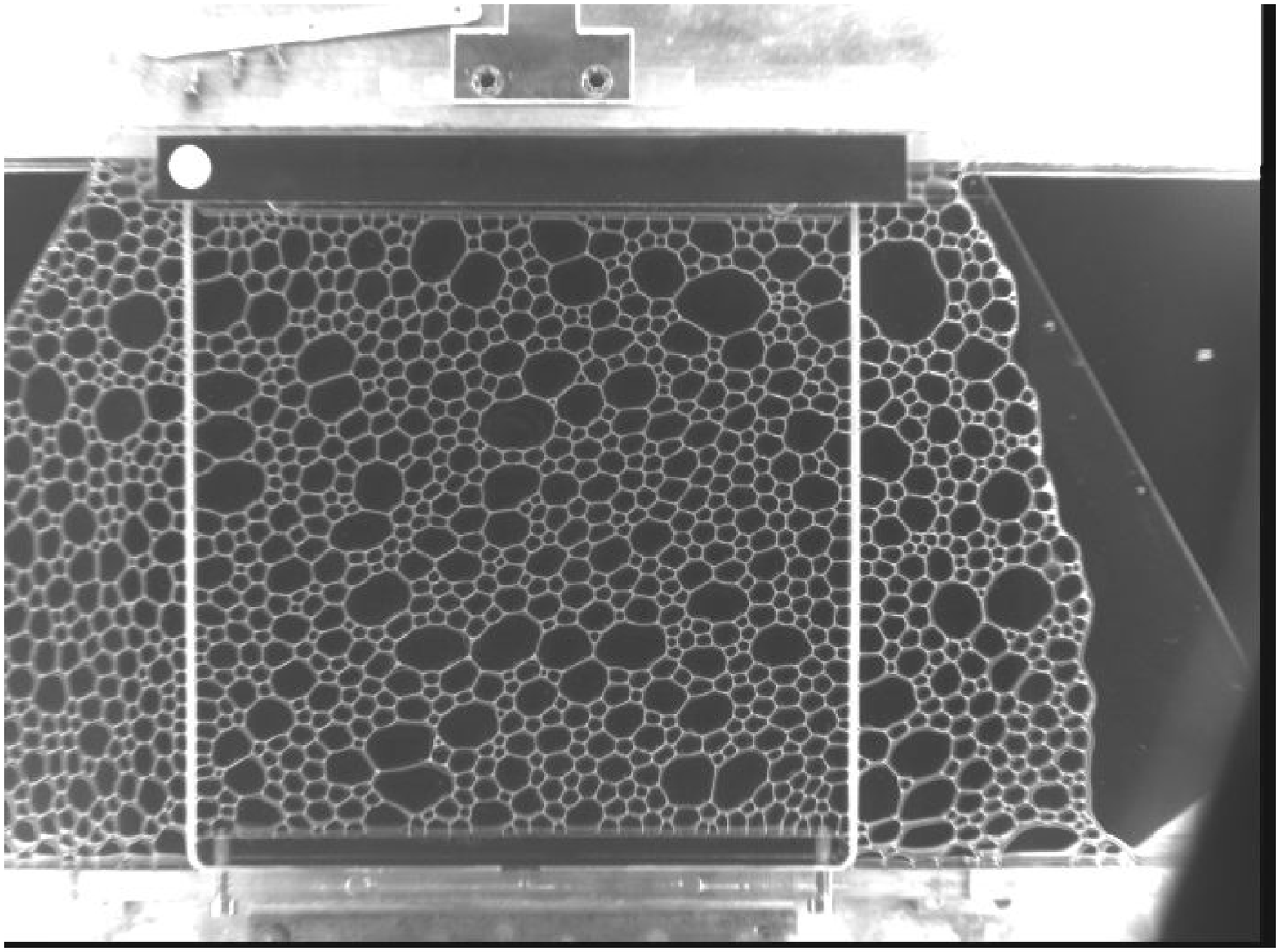}
{\small c)}
 \includegraphics[width=2.7cm]{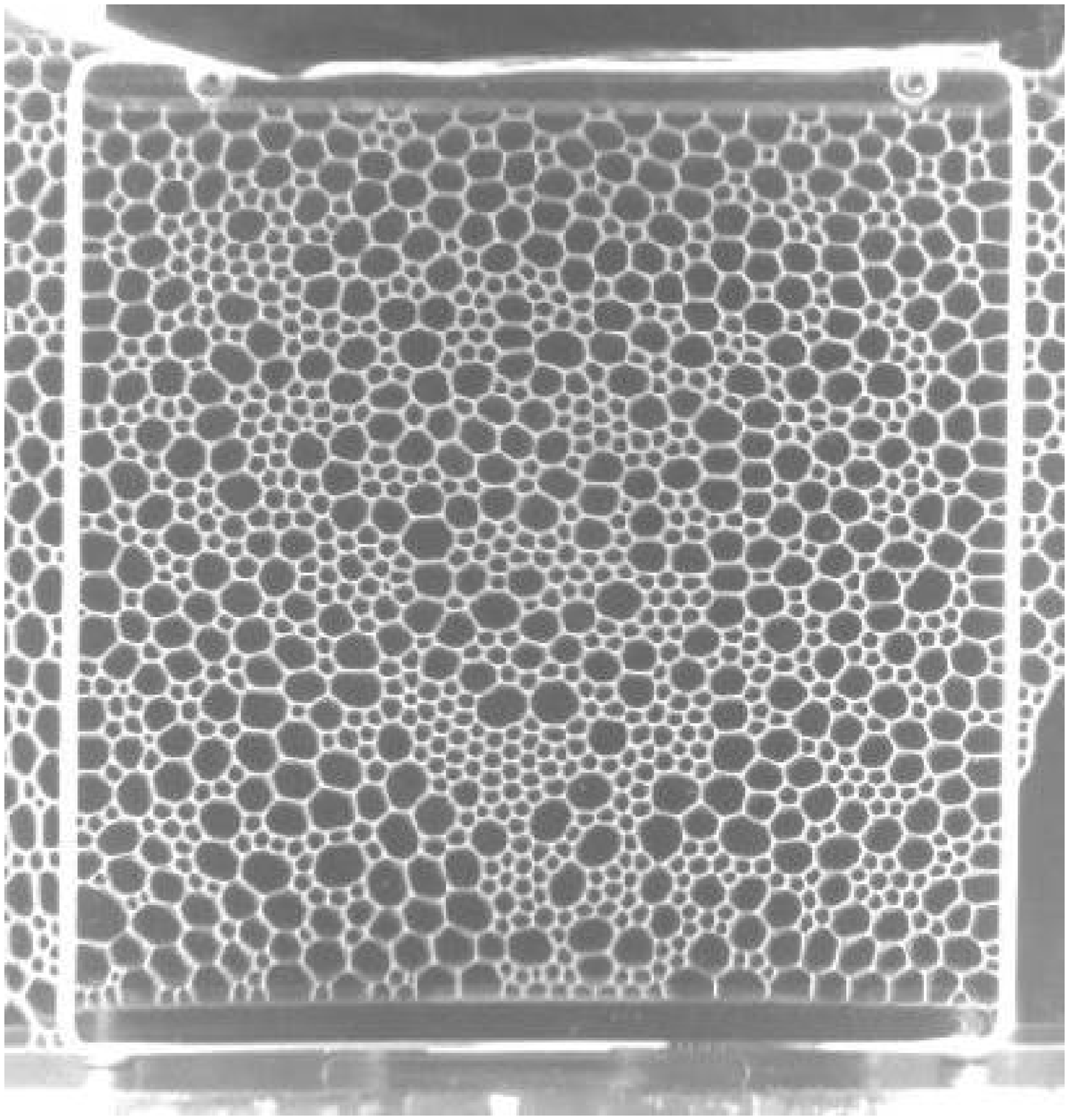}
 \\
{\small d)}
\includegraphics[width=4.7cm]{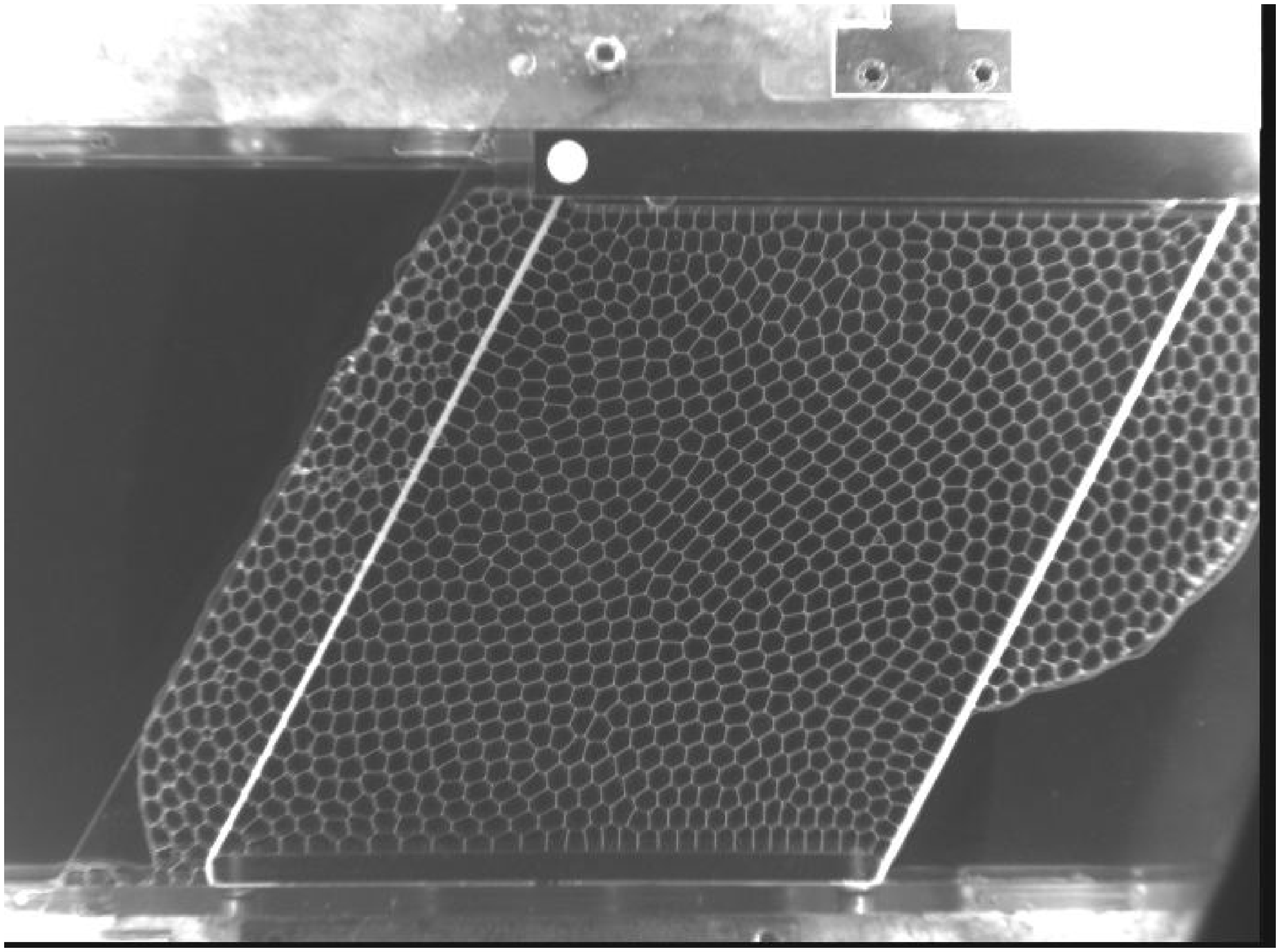}
{\small e)}
\includegraphics[width=4.7cm]{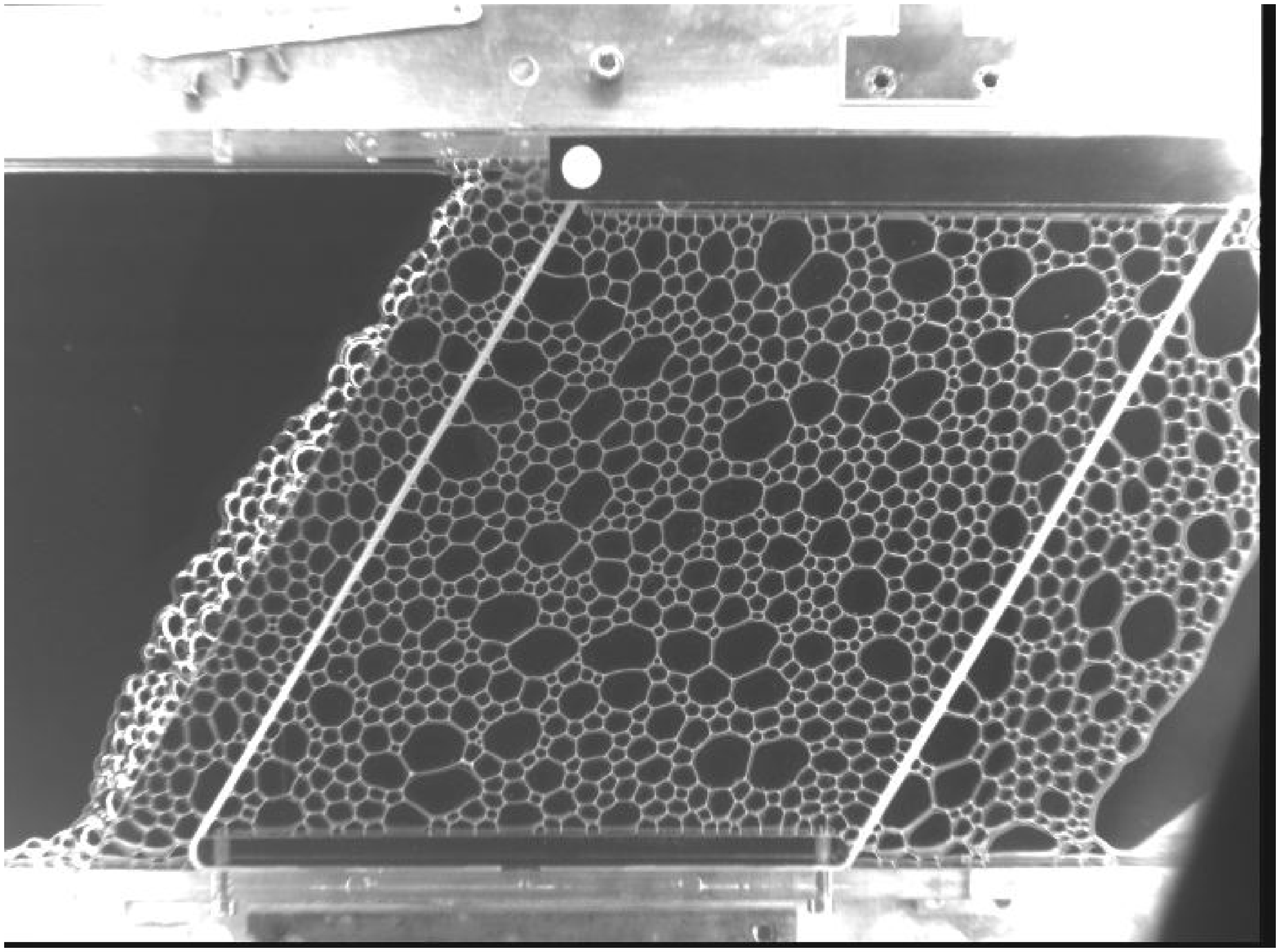}
{\small f)}
 \includegraphics[width=2.7cm]{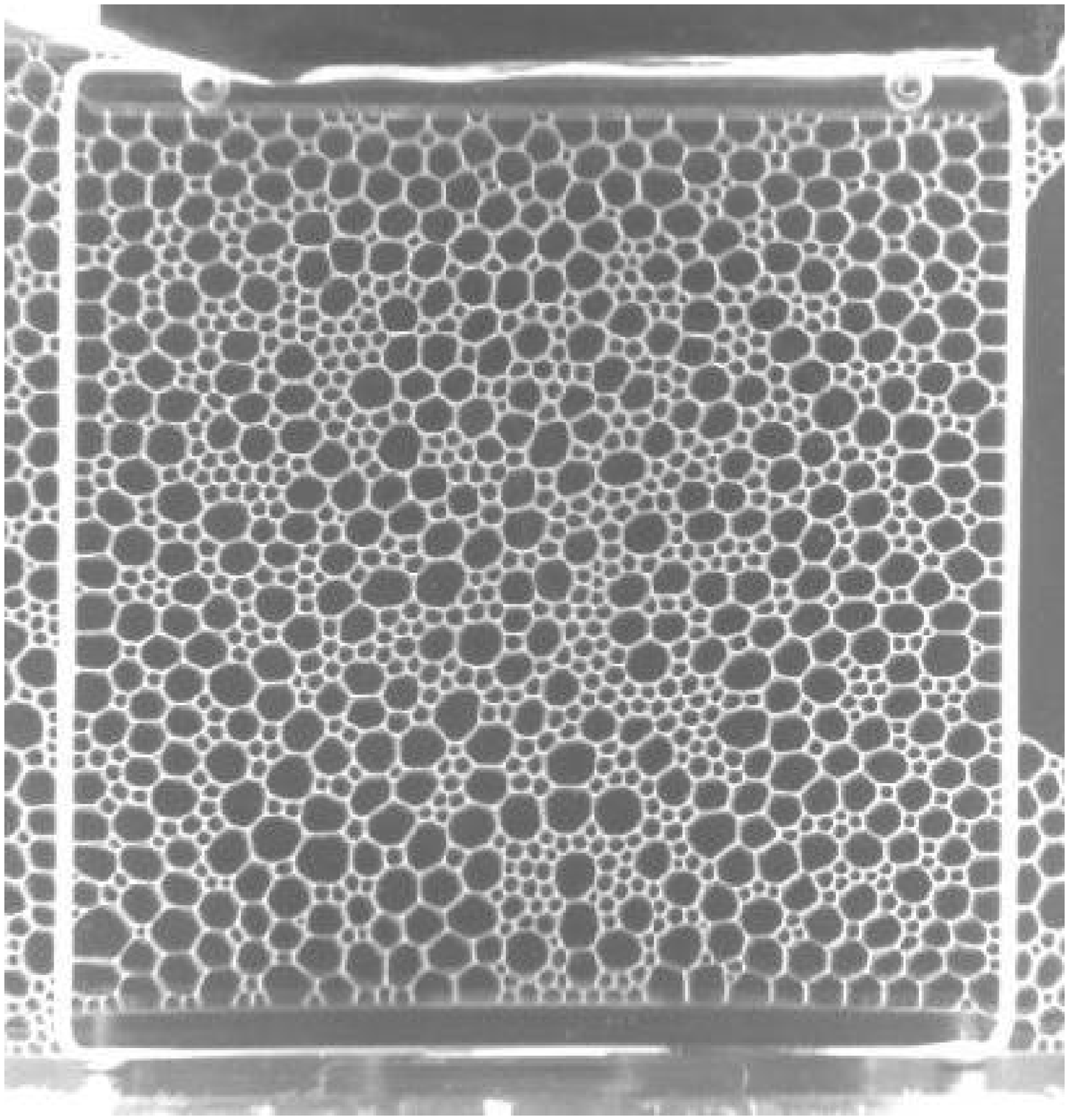}
\caption{The foam is initially prepared within a square of 18 cm side (a,b,c), which can be deformed into a parallogram of the same area, here at an angle of 30$^\circ$ (d,e). The images show samples with the smallest (a,d) and largest (b,e) width $\Delta A/ \langle A \rangle$ of the bubble area distribution.
(f) Same bimodal foam as in (c), after 44 cycles of shear.
}
\label{shear}
\end{figure}

The trough is a rectangle 44 cm long and 20 cm wide. The foam is enclosed in a smaller square of size $18 \times 18=324$  cm$^2$, with four independent boundaries (Fig. \ref{setup}b). A rigid rod (fixed boundary) is attached to the trough itself. Another rigid rod (moving boundary) is attached to the glass lid, which slides laterally under teh control of a motor. The rectangle is closed on each side by a rubber band (passive boundaries): it thus deforms into a parallelogram at constant area (Fig. \ref{shear}).

\begin{figure}
		{\small (a)} \hspace{3cm}
		{\small (b)}\hspace{3cm}
		{\small  (c)}\hspace{3cm}
		{\small (d)}\\
		\includegraphics[width=3.3cm]{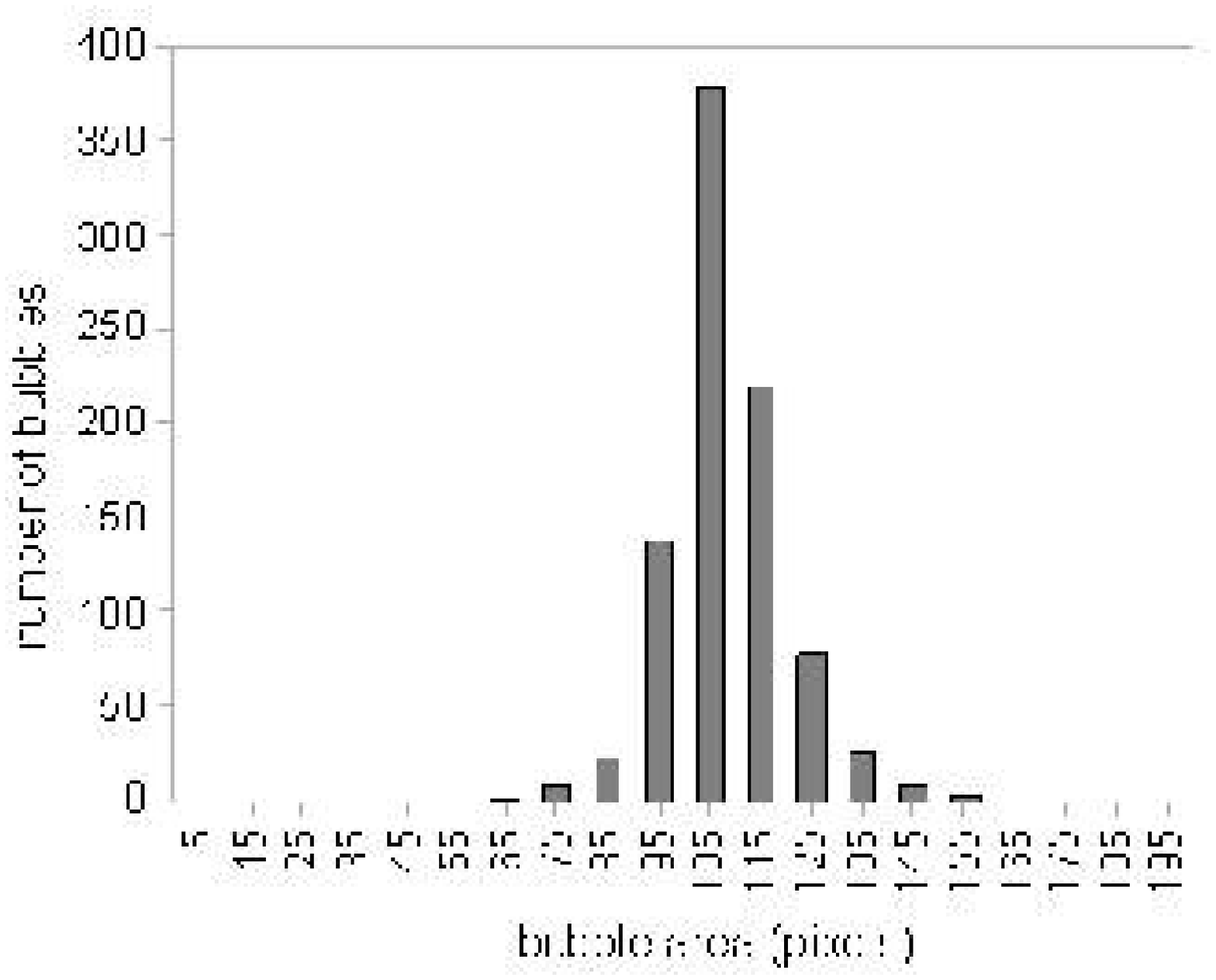}\hspace{-0.3cm}
		\includegraphics[width=3.3cm]{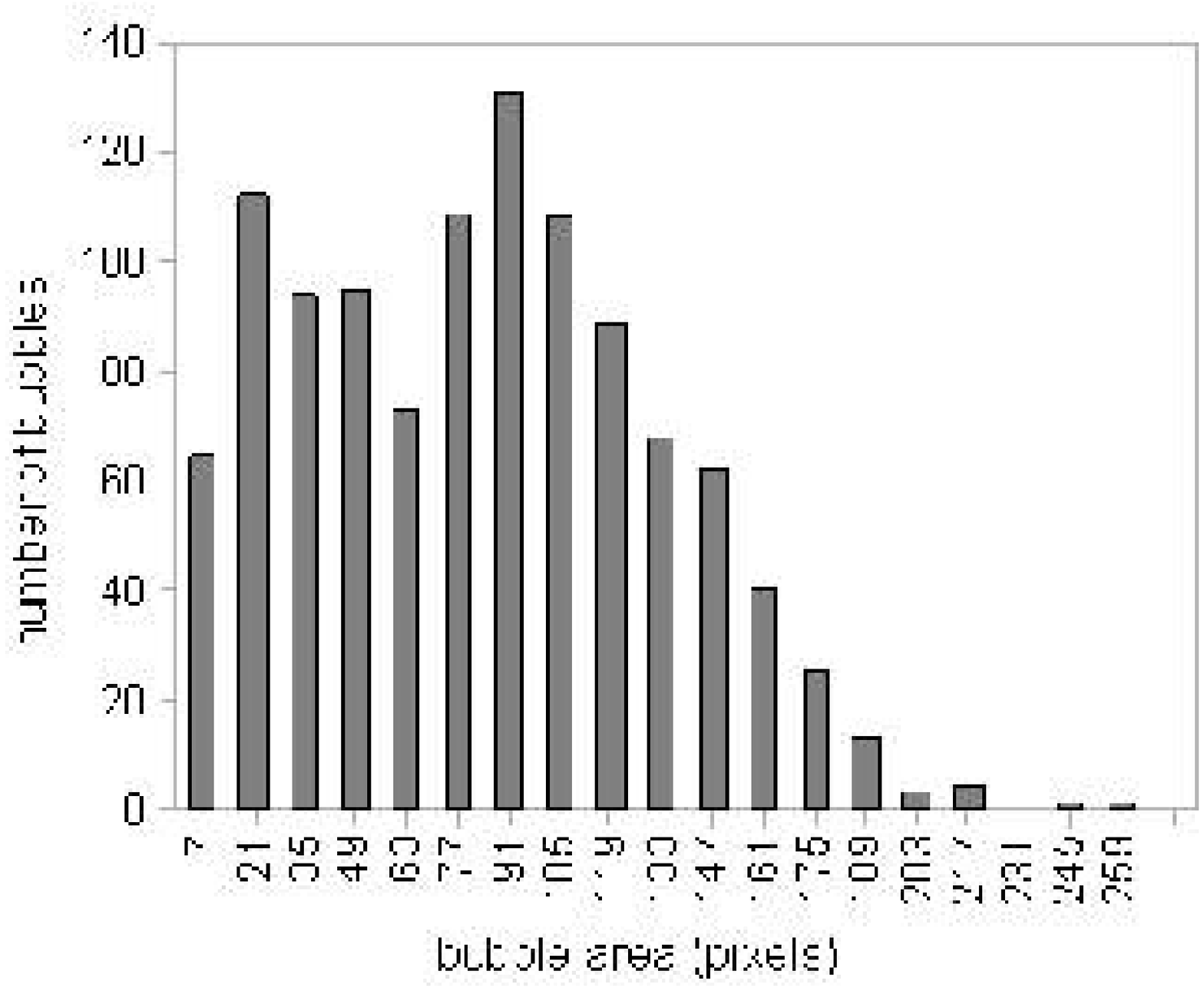}\hspace{-0.3cm}
		\includegraphics[width=3.3cm]{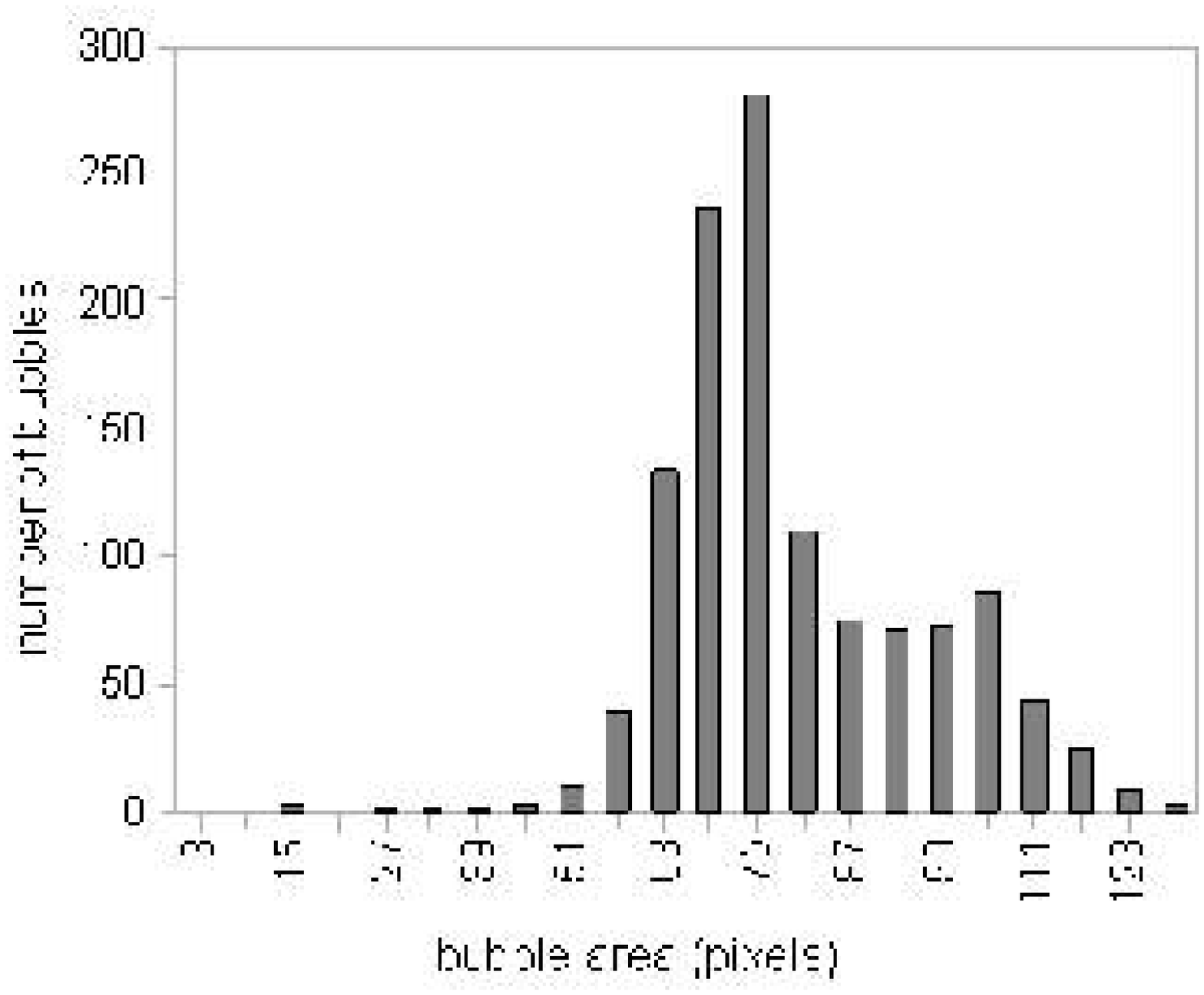}\hspace{-0.3cm}
		\includegraphics[width=3.3cm]{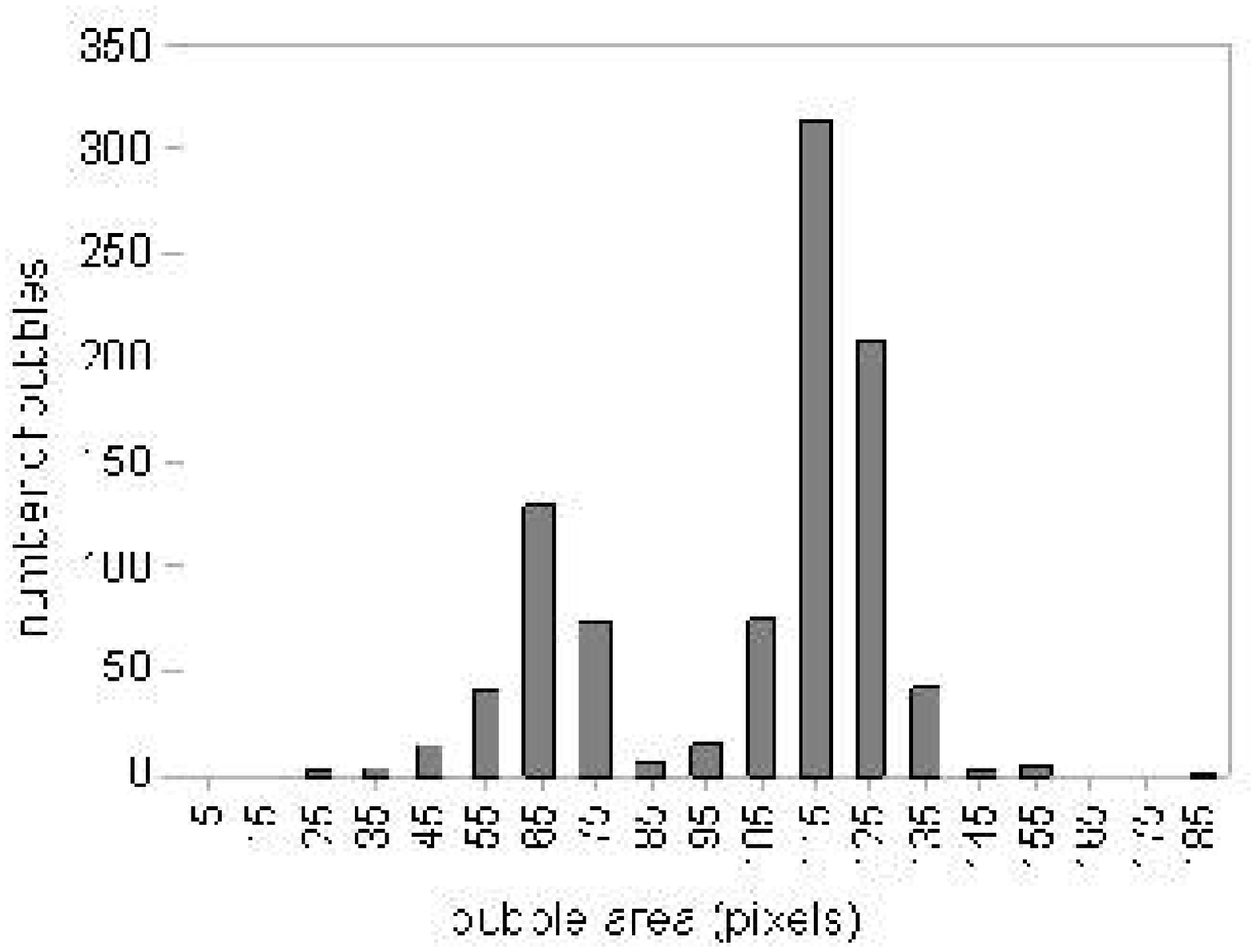}
		\includegraphics[width=3.3cm]{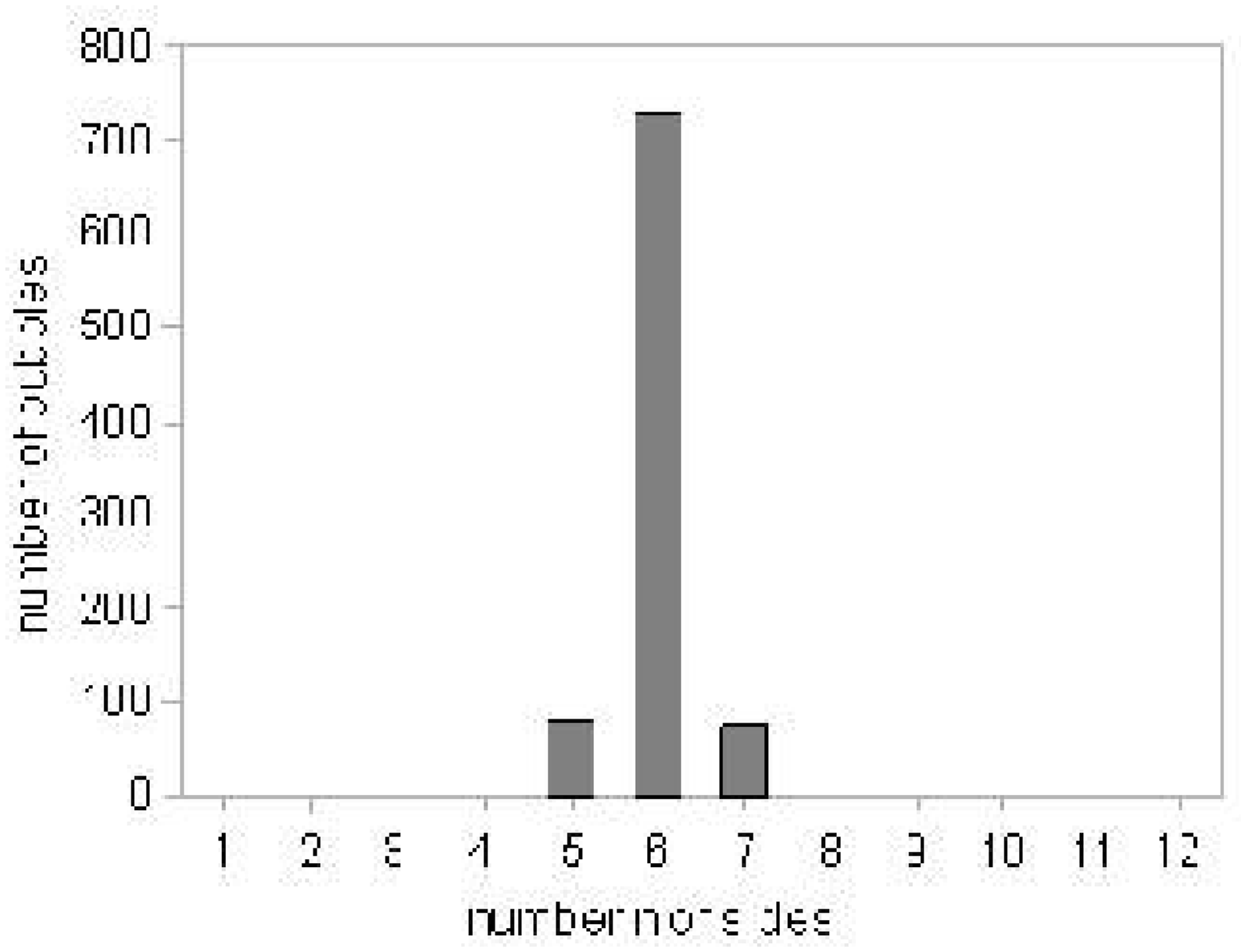}
		\includegraphics[width=3.3cm]{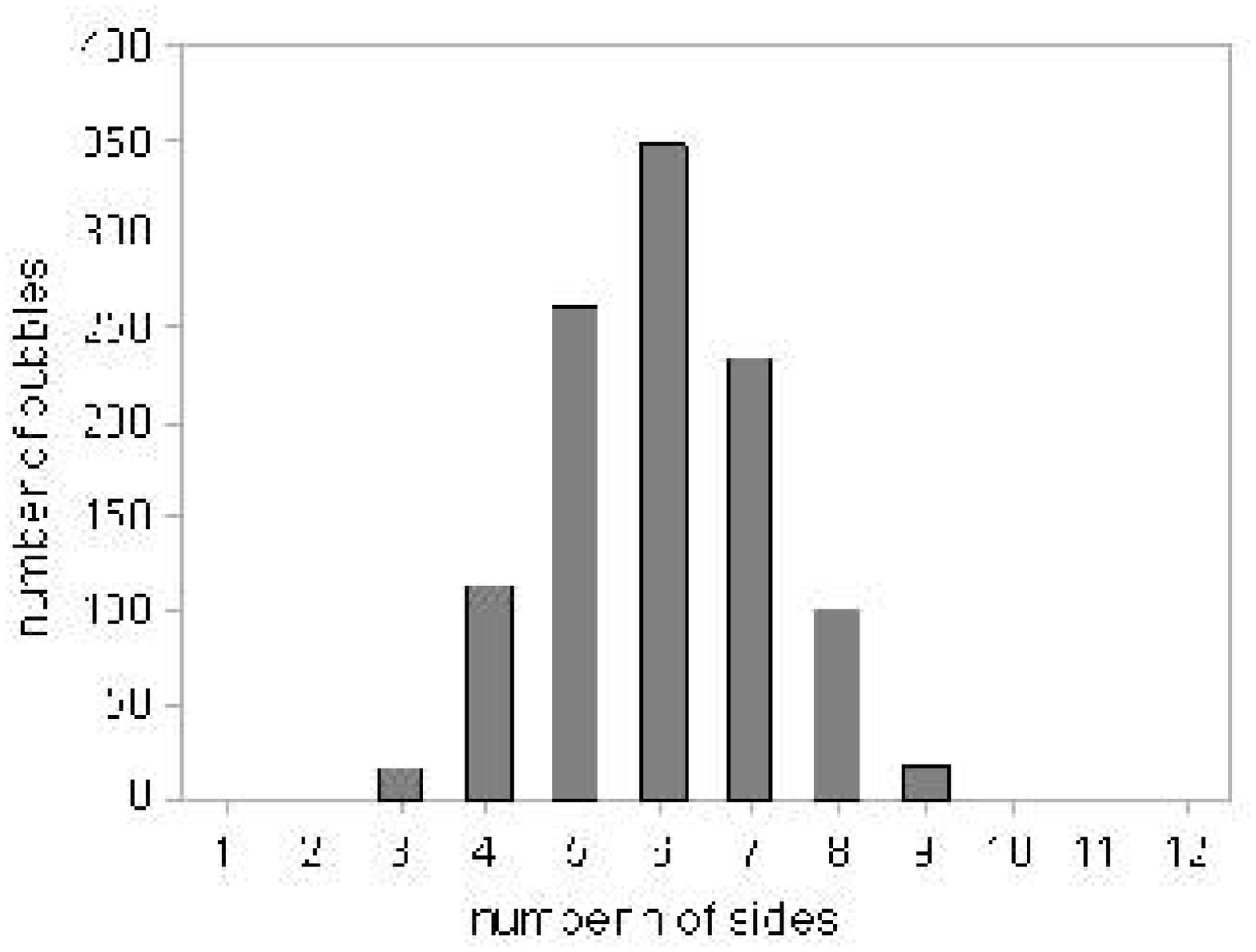} 
		\includegraphics[width=3.3cm]{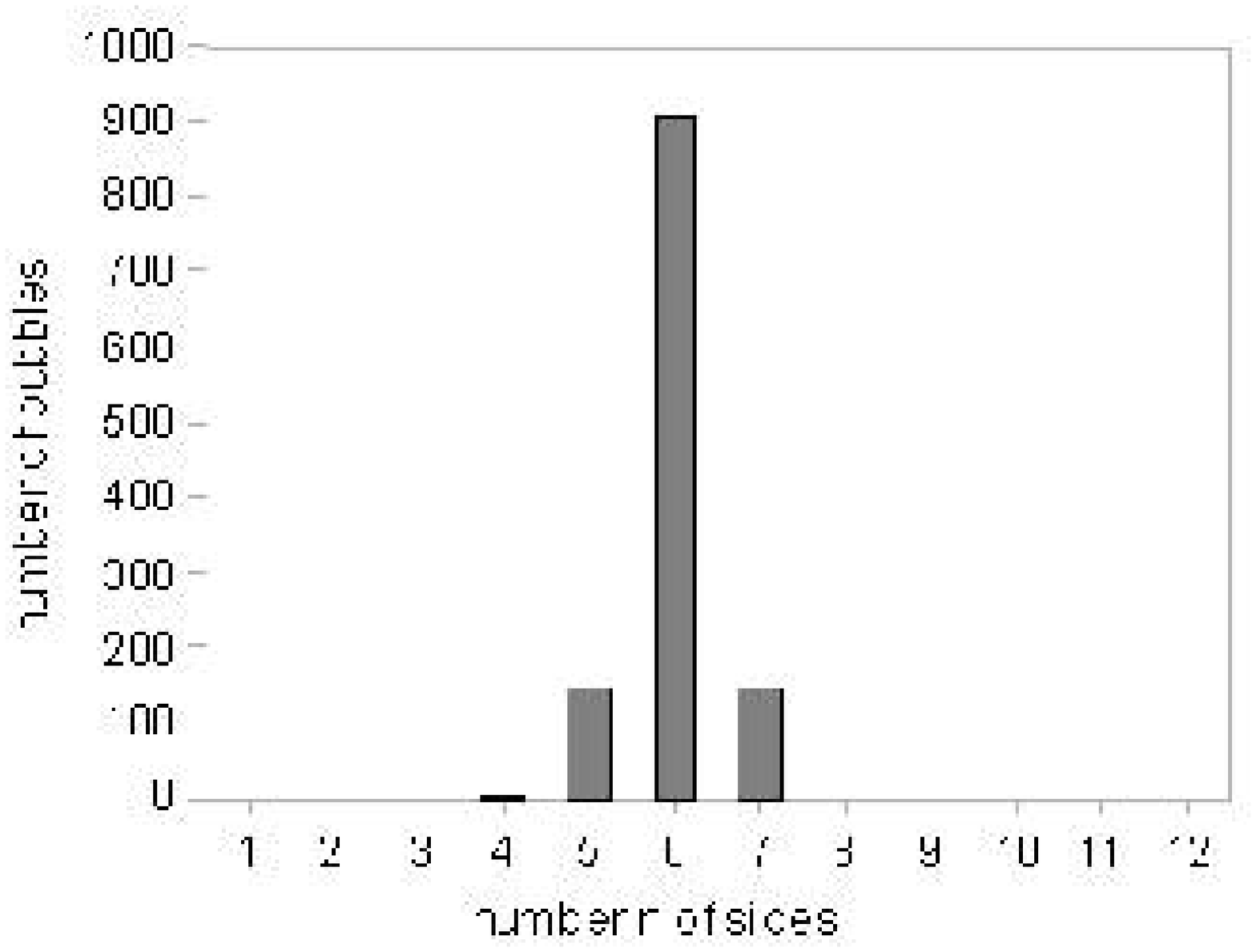} 
		\includegraphics[width=3.3cm]{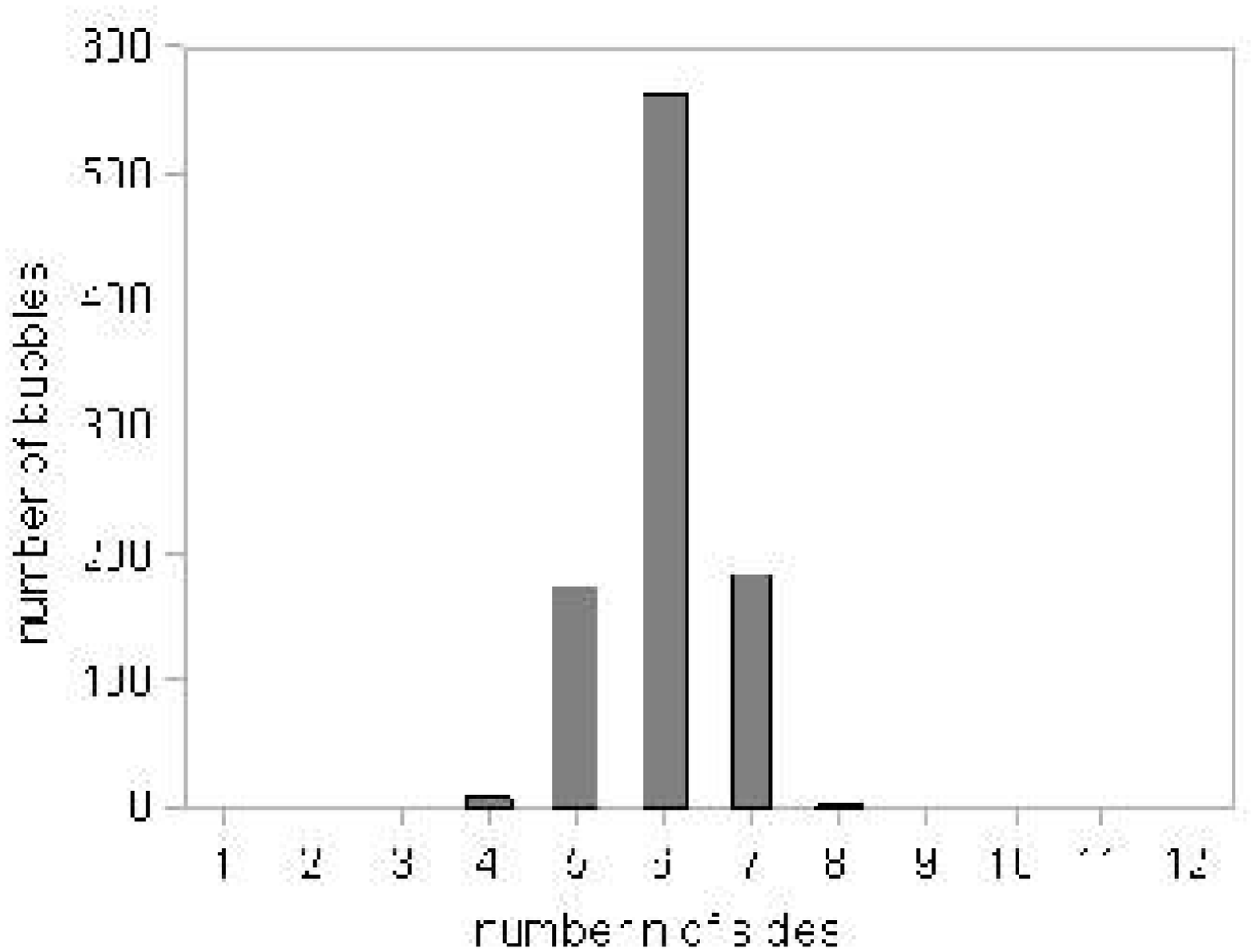}
\caption{ 
Four examples of histograms of bubble area $A$ (top) and number $n$ of sides (bottom) in freshly prepared foams:
(a) narrow monomodal, (b) polydisperse, 
(c) with shoulder, (d) bimodal.}
\label{histos}
\end{figure}

The trough is filled with a solution of 10\% commercial dishwashing liquid in water. A pump blows air into a tube placed below the surface of water, which is moved back and forth to produce an homogeneous foam. At low flow rate, the solution surface tension and the tube diameter together fix the bubble volumes \cite{clift}. The histogram of bubble volumes has thus a single peak. Its width  increases with the flow rate,  from ``narrow monomodal" (Figs. \ref{histos}a) to  ``polydisperse" (Figs. \ref{histos}b). Using two tubes  simultaneously yields a histogram of bubble volumes with two peaks.
The distance between peaks can be chosen smaller than their width (``with shoulder", Fig. \ref{histos}c), or larger (``bimodal", Figs. \ref{histos}d).

The foam thickness $h$  is the distance between water and glass in the absence of bubbles (Fig. \ref{setup}a). At small $h$ the foam is wet, with round bubbles. At large $h$ bubbles begin to pile on top of each other making a 3 dimensional foam \cite{coxwv01}. In between ($3 < h< 6,8$ mm),  the bubble monolayer is moderately wet; bubbles can deform and thus store elastic energy. The foam has a finite shear modulus and thus behaves as a 2D elastic solid until it yields and bubbles rearrange. 
 
\subsection{Measurements}

A circular fluorescent tube (diameter 48 cm) is placed just above the trough. A video camera records a 33 cm wide, 25 cm high field of view. Images are analysed using ImageJ \cite{imageJ} and skeletonised: white bubbles are surrounded by black boundaries one pixel thick. 
Each bubble's number $n$ of sides ({\it i.e.} of neighbours)  is  measured without ambiguity.  At a given $h$, each bubble's volume determines its projected area seen from the top \cite{yield}.
Hereafter, we call ``bubble area" the area $A$ of each skeletonized bubble (this corresponds, in the original image, to the area of the gas and almost all the liquid). 

For each foam, we plot the histogram of $A$ and  $n$; Fig. \ref{histos} shows different examples. We check that
both  averages,  $\left \langle A \right \rangle$ and  $\left \langle n \right \rangle$, have the expected values.
 Here  the total number $N$  of bubbles  is typically $570$ to $1800$ 
(with a maximum of $2719$); $\left \langle A \right \rangle$ is close to the total area of the foam divided by  $N$  (that is, the inverse of the bubble density), typically $18$ to $65$ mm$^2$;  $\left \langle n \right \rangle$ is always equal to 6, minus a small correction of order $1/N$  \cite{graustein,weairerivier}.
 
  We then measure the  standard deviations 
 $\Delta A
  = \left(\left \langle A^2 \right \rangle - \left \langle A \right \rangle^2\right)^{1/2}$ and 
  $\Delta n = \left(\left \langle n^2 \right \rangle - \left \langle n \right \rangle^2\right)^{1/2}$. 
  To enable  comparisons, in what follows we only consider the dimensionless standard deviations
   $\Delta A/\left \langle A \right \rangle$ and $\Delta n/\left \langle n \right \rangle$ as measurements of the geometrical and topological disorder:
\begin{equation}
\frac{\Delta A}{\left \langle A \right \rangle} 
= \sqrt{\frac{\left \langle A^2 \right \rangle}{\left \langle A \right \rangle^2} -1 } 
\quad \quad ; \quad \quad
\frac{\Delta n}{\left \langle n \right \rangle} 
= \sqrt{\frac{\left \langle n^2 \right \rangle}{\left \langle n \right \rangle^2} -1 } 
. \label{Deltas}
\end{equation}
In principle it is possible to define $n$ unambiguously even for bubbles at the boundaries
 \cite{granerjjf01}. While useful in simulations,  in experiments the image analysis and skeletonization   
 cause some systematic bias in the area measurement at the boundaries.
 We thus choose to remove these bubbles when evaluating both measures of disorder, which are very sensitive to artefacts, or to errors due to even a small number of bubbles \cite{glazier}.
 
\subsection{Shear cycles}

A shear cycle consists in shearing the foam by displacing the movable boundary a distance 10.4 cm 
and back, and then to $-10.4$ cm, 
and back again. This corresponds to an angle of $+ 30^\circ$ to  $- 30^\circ$
and  a strain amplitude of $+ 10.4/18 = + 0.58$ to $-0.58$.
For dry foams, to be well above the yield strain requires a larger amplitude 
  \cite{rau07}; we obtain  it by  preparing the foam when the moveable rod is at the position  $-4.1$ cm (at the left limit of the camera field of view): the cycle is then at positive strain from 0 to  
   $(10.4+4.1)/18=0.8$, then down to $- (10.4-4.1)/18=-0.35$.
  For the driest foam, we also reduce the size of the trough by reducing the distace between the fixed and moveable rods to 10.5 cm; $N$ is then smaller (413 {\em vs.} 879 bubbles), and a cycle consists of 
  positive strains in the range from 0 to  $(10.4+4.1)/10.5=1.4$, then down to $- (10.4-4.1)/10.5=-0.6$.
   
A cycle lasts $T=275$ seconds and the maximum velocity is $37.9^\circ/$min, slow enough that the
results presented here do not depend on the velocity.
The bubble deformation and the velocity gradient enforced by the lateral boundaries are uniform
 (up to non-affine displacements due to  local disorder). 
During shear, each cell area is conserved, so that $\Delta A/\left \langle A \right \rangle$ keeps its initial value. We measure $\Delta n/\left \langle n \right \rangle$ at each period.

\section{Results}
   
 
Figure \ref{shear} shows foam samples with the smallest (Fig. \ref{shear}a,d) and largest (Fig. \ref{shear}b,e) width $\Delta A/ \langle A \rangle$ of the bubble area distribution. We  make the different kinds of distribution overlap, {\it i.e.} the width of some polydisperse foams are larger than some bimodal foams.

In principle, it would be possible to prepare foams with a slightly larger  width of the area distribution by including a few very large bubbles. However, this would make the distribution  spatially heterogeneous.
At the other extreme, a foam with a smaller width of the area distribution would be possible with other experimental set-ups, but with only a small  number of non-hexagonal bubbles, the distribution of the number of sides would again become spatially heterogeneous (see caption of Fig. \ref{cath_dav}).

		
\begin{figure}
		{\small (a)} \hspace{4cm}
		{\small (b)}\hspace{4cm}
		{\small  (c)}\\
\includegraphics[width=4.5cm]{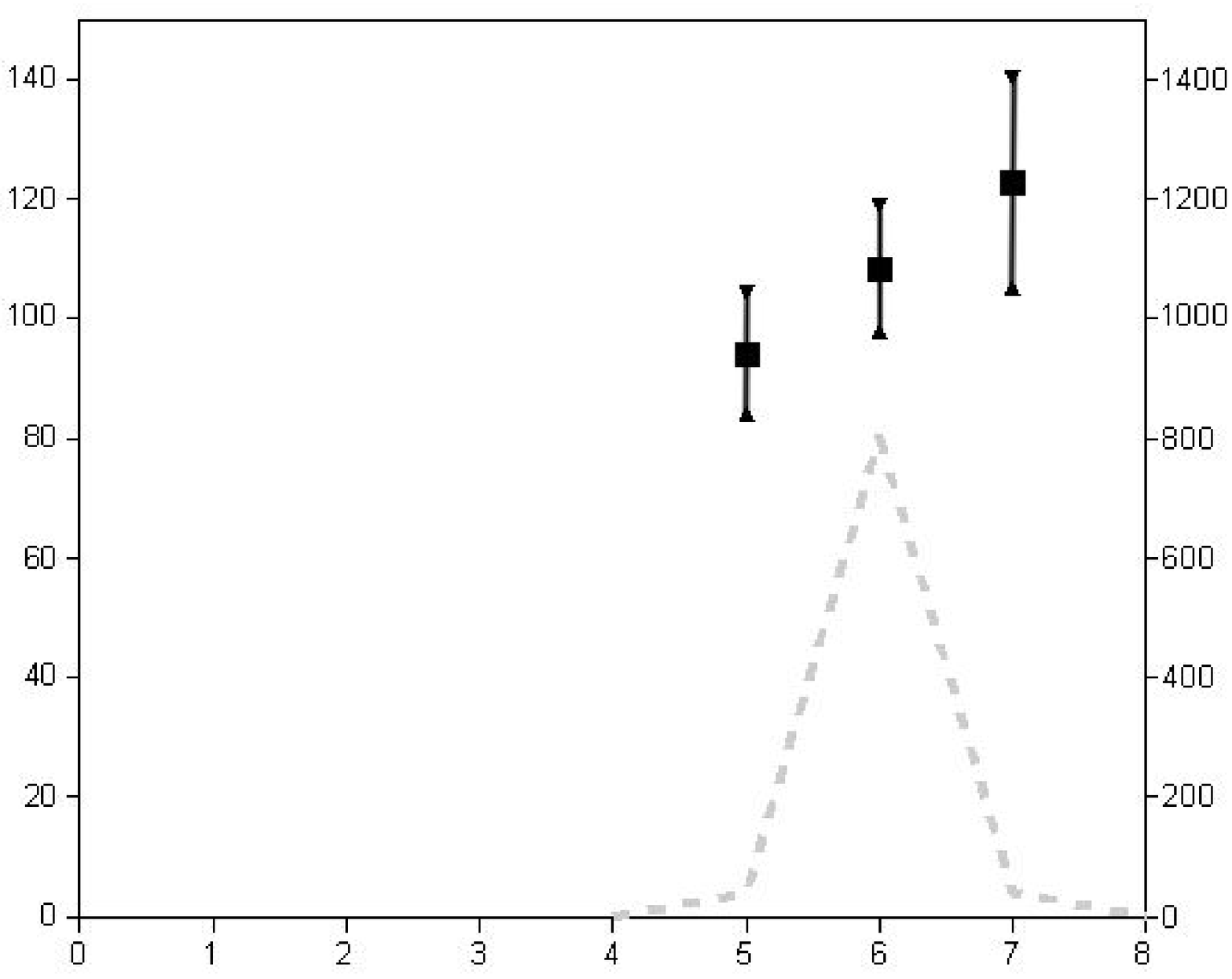}
\includegraphics[width=4.5cm]{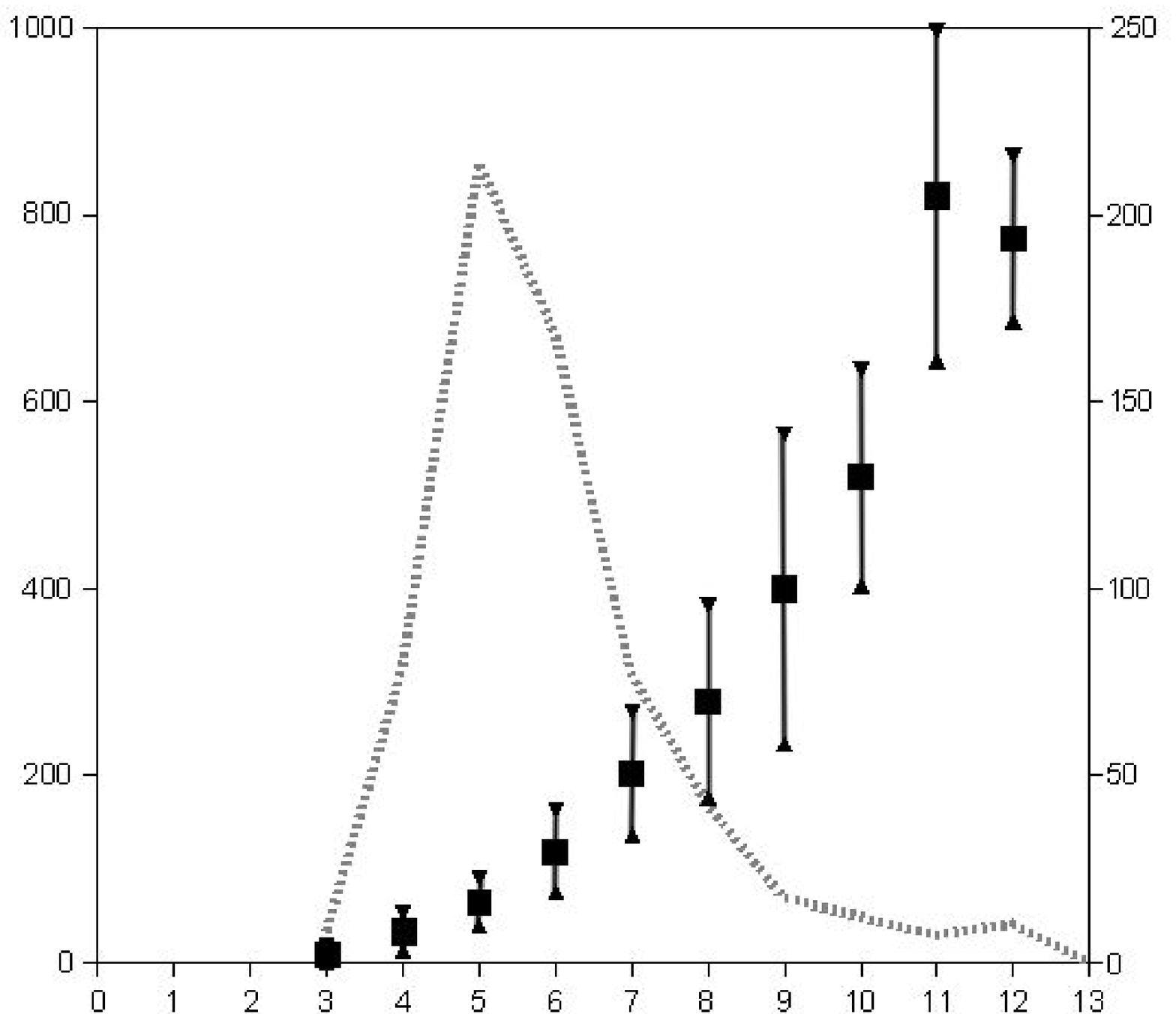}
\includegraphics[width=4.5cm]{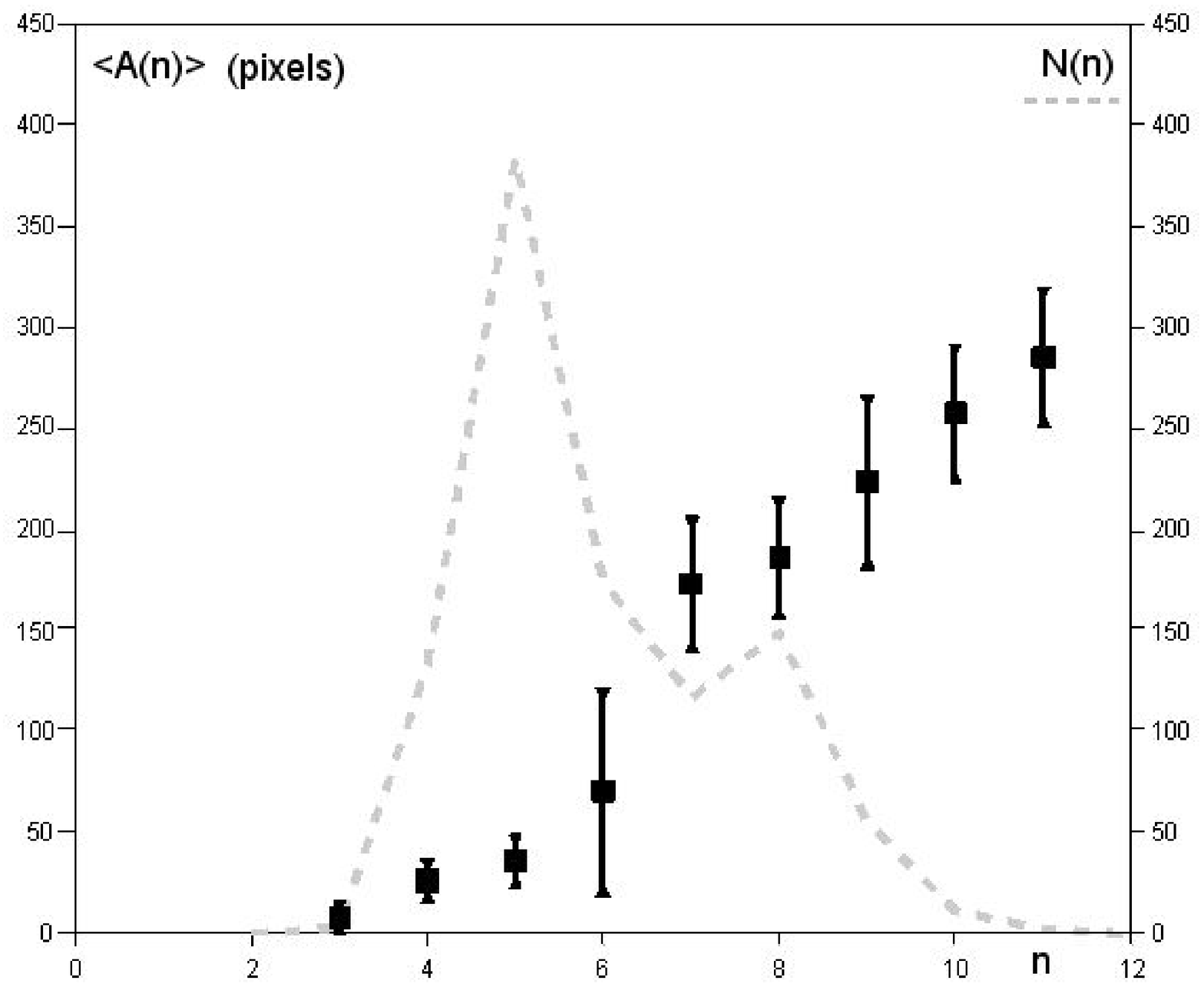}
\caption{ 
Size-topology correlation for foam experiments after several shear cycles: (a) narrow monomodal, (b) polydisperse, (c) bimodal.
Closed squares, left scale: average size $<A(n)>$  (in pixels) of the population of bubbles with  $n$ sides; bar:  the standard deviation calculated on each population.
Dashed line, right scale:  number $ N(n)$  of bubbles in this population.
}
\label{lewis_graph}
\end{figure}

Fig. \ref{lewis_graph} indicates a clear correlation between bubble size and number of sides
for sheared foams (unsheared foams are similar, data not shown). 
There is still debate (for reviews see \cite{weairerivier,glazier,lewis,smith,lisso,pina,cargese,bideau,fortes}) concerning (i) systems for which this relation is or is not linear; (ii) whether it is related to entropy maximisation, energy minimisation, or both (free energy minimisation); (iii) what is the value and significance of the linear intercept; (iv) whether $n$ correlates better with bubble perimeter or area (or volume, in 3D); (v) and especially up to what precision the relation can significantly be considered as linear for practical or theoretical applications.
Here, in a narrow monomodal foam there are too few points to reach a conclusion; in  polydisperse and bimodal foams the relation is certainly not linear. 


\begin{figure}
		{\small (a)} \hspace{6cm}
		{\small (b)}\\
 \includegraphics[width=6cm]{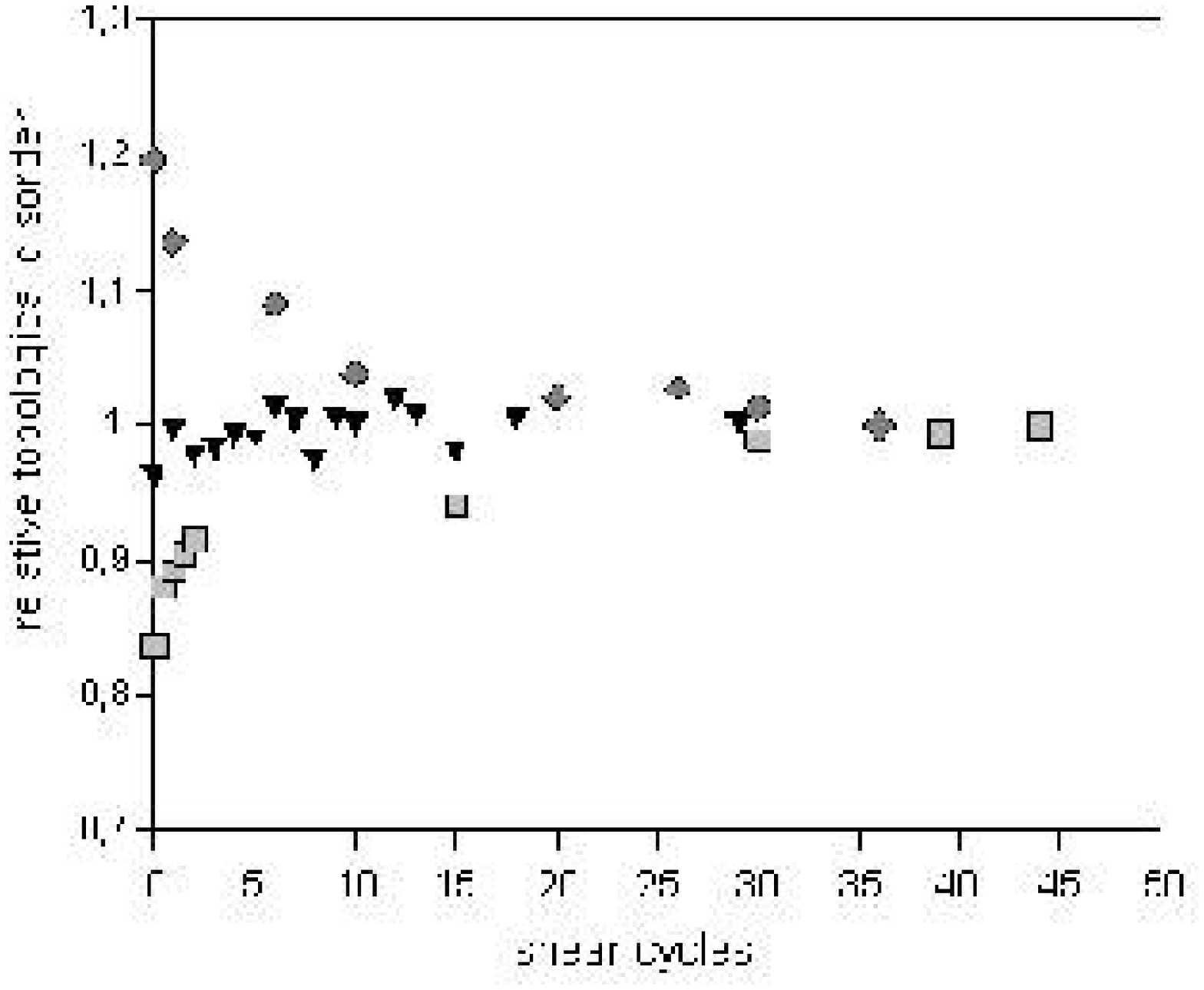}
\includegraphics[width=6cm]{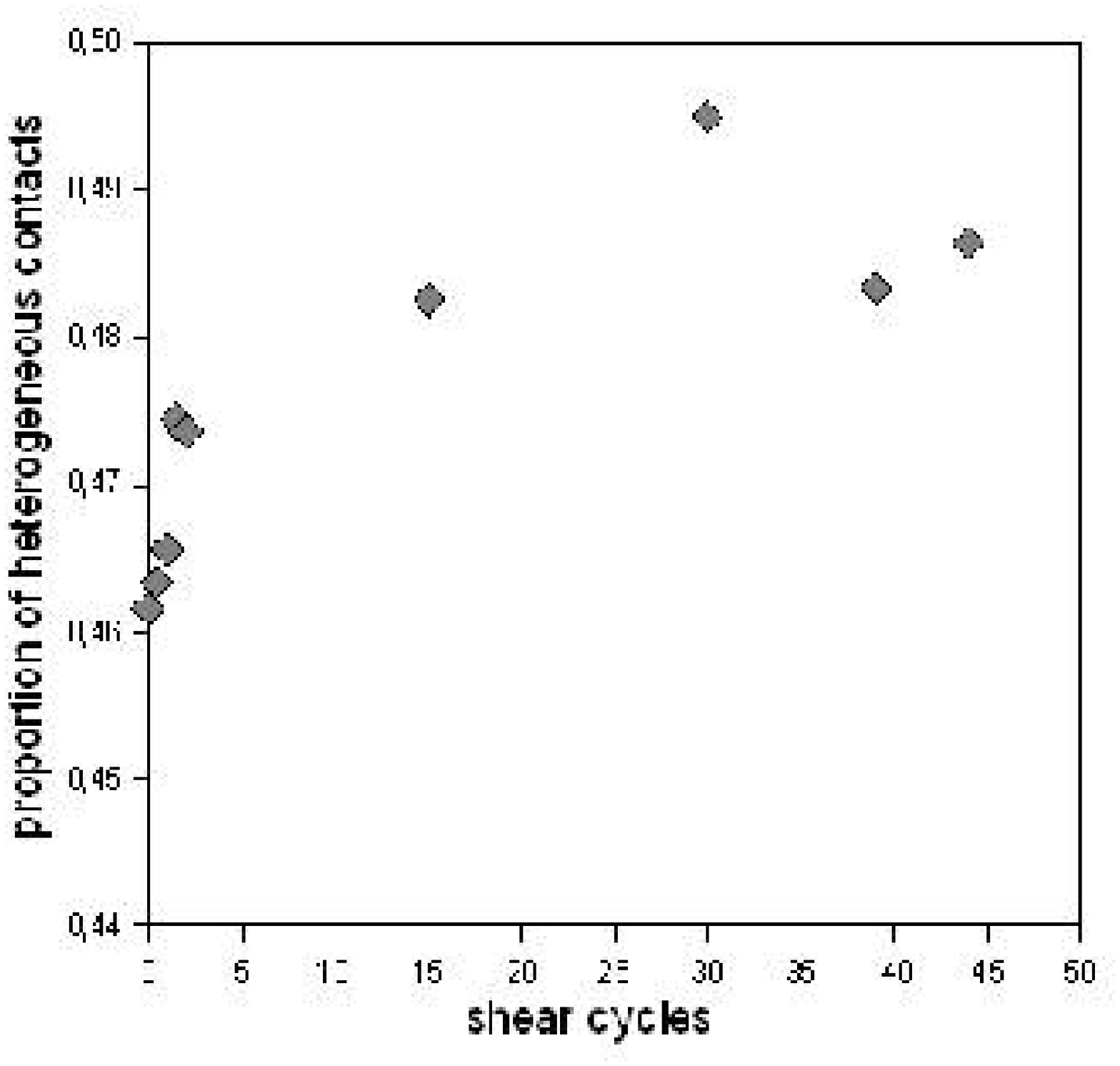}
\caption{ 
(a) 
Evolution  of topological disorder under shear for given bubble area distribution:
 $\Delta n/ \langle n \rangle$ (normalised by its asymptotic value to facilitate comparisons) {\it versus}  number of shear cycles. Data correspond to Fig.  \ref{lewis_graph}: diamonds: narrow monomodal; triangles: polydisperse; squares: bimodal.
(b) Evolution of the fraction of contacts between unlike (small$-$large) bubbles during shear for a bimodal foam. 
 }
\label{Deltan_vs_t}
\end{figure}
 
  While the foam is sheared, the bubbles rearrange and the distribution of the number of sides changes while the area distribution remains fixed.  In all experiments, $\Delta n/ \langle n \rangle$ reaches  a constant value  after a few shear cycles (Fig. \ref{Deltan_vs_t}a).
 Under shearing, weakly disordered foams become more ordered: $\Delta n/ \langle n \rangle$ decreases; bimodal foams (Fig. \ref{shear}c,f) become increasingly mixed 
 (Fig. \ref{Deltan_vs_t}b), so that $\Delta n/ \langle n \rangle$ increases. 
 This is important, and not trivial \cite{teixeira}.

 
\begin{figure}
\begin{center}
\includegraphics[width=8cm]{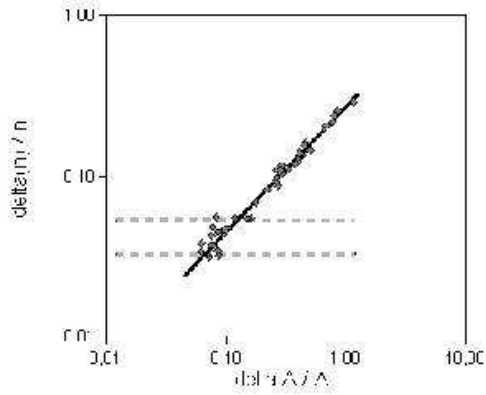}
\end{center}
\caption{
Correlation between measures of disorder in sheared foams.
Each point is the asymptotic value of $\Delta n/ \langle n \rangle$ and $\Delta A/ \langle A \rangle$, from Fig. (\ref{Deltan_vs_t}a).
Solid line: power law    $\Delta n/ \langle n  \rangle = 0.27 \; (\Delta A/ \langle A \rangle)^{0.8}$.
Dashed lines: lower bound for topological disorder, for the extreme values of 
$N$ in these experiments (413 and 2719).
When enclosing a honeycomb in a square box, boundary conditions impose  at least two grain boundaries with length of order $N^{1/2}$  between perpendicular lattices. 
Topological considerations \cite{Alfonso} imply a minimal number of paired five- and seven-sided bubbles per unit line of grain boundaries. We then estimate the topological disorder to be 
 at least  $1 / [3 \times 12^{1/8} \times N^{1/4}].$
}
\label{cath_dav}
\end{figure}
  
Any given pattern can be represented by a point  in the plane  $\left(\Delta A/ \langle A \rangle,\Delta n/ \langle n \rangle\right)$. During a single shear experiment, this point moves up or down. 
After many cycles, points  tend to cluster on a single curve (Fig. \ref{cath_dav}), namely the power law:
\begin{equation}
\frac{\Delta n}{\langle n \rangle} =
0.27 \;  \left(\frac{ \Delta A}{ \langle A \rangle}\right)^{0.8}.
\label{fit_ecart_type}
\end{equation}
This is the main result of this paper.

\begin{figure}
(a) 
\includegraphics[width=6cm]{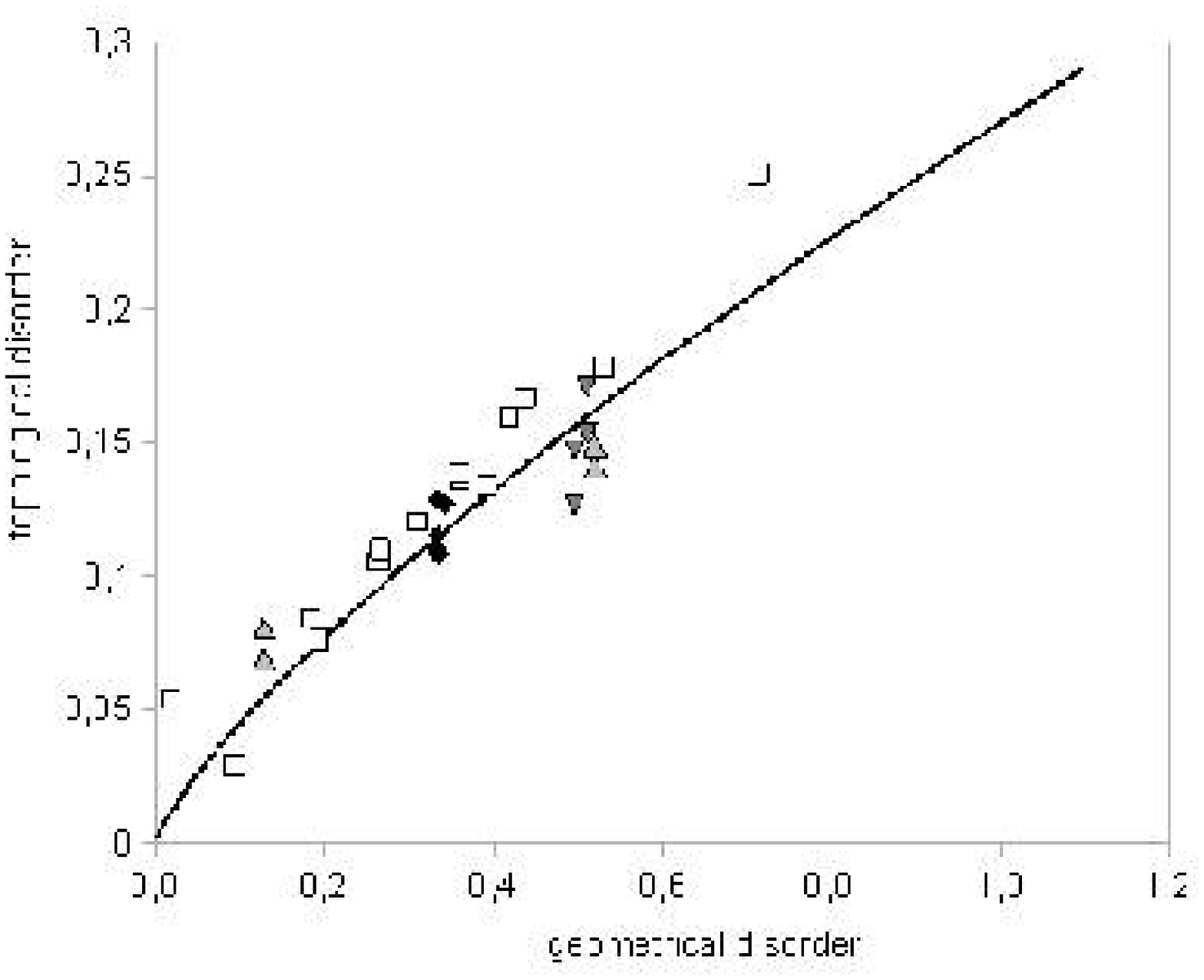}
(b) 
\includegraphics[width=6cm]{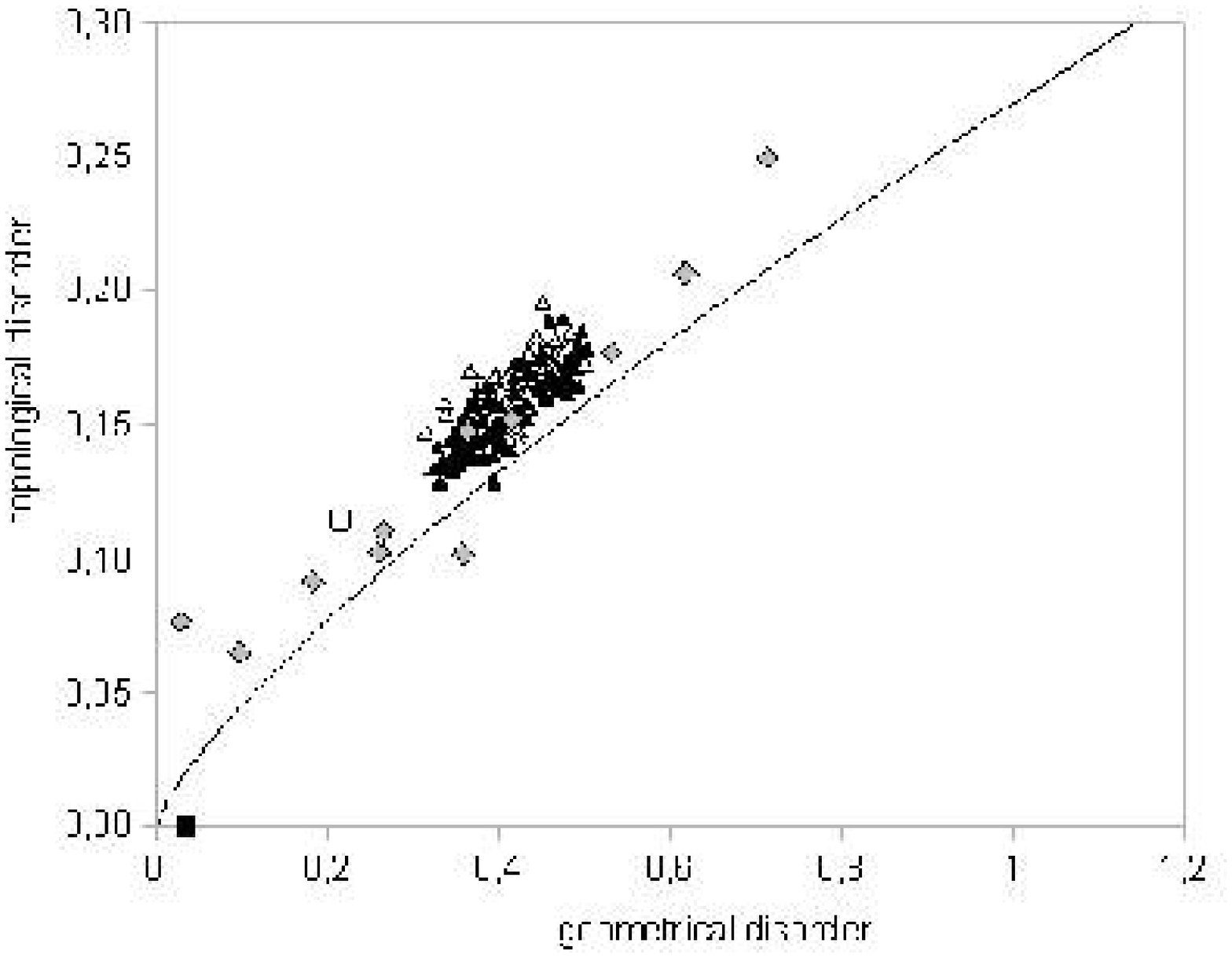}
\caption{ 
Correlation in disorder for other patterns. Solid line: eq. (\ref{fit_ecart_type}), same as in Fig. \ref{cath_dav}.
(a) Foam simulations: downward-pointing triangles, Surface Evolver \cite{brakke} simulations \cite{rau07} shuffled by cycles of shear, with four boundaries as in the present experiments; upward-pointing triangles, same, but with periodic boundary conditions (no boundaries)  \cite{rau07}; closed diamonds, Surface Evolver simulations shuffled by cycles of uniaxial elongation \cite{simon_elong}; open squares, simulations using the Potts model at constant tension, with periodic boundary conditions, shuffled by increasing the fluctuations  \cite{Maree.scbm07,jos_pnas}. 
(b) Biological tissues: closed triangles, cells rearranging during the formation  of a fruit fly (Drosophila) embryo  thorax epithelium; open triangles, same for {\it mud} mutant tissue  \cite{courtypreprint};
closed square, 
facets (ommatidia) in a  fruit fly (Drosophila) retina are arranged in an ordered honeycomb;
open square, same for the {\it rough eye} mutant retina.
\cite{Hayashi.n04}; grey diamonds, simulations using the Potts model of cells with adhesion and cortical tension, with periodic boundary conditions, shuffled by increasing the fluctuations 
 \cite{Maree.scbm07,jos_pnas}. 
}
\label{fig:complete}
\end{figure}
    
\section{Discussion}

In our foams, the width of the area distribution varies from 0.1 to 1, and that of the topology distribution from 0.03 to 0.3. In this limited range,
eq.  (\ref{fit_ecart_type}) means that $\Delta n/ \langle n  \rangle $ is close to $0.3 \; \Delta A/ \langle A \rangle$.  Distributions with the same width but different higher-order moments  behave similarly. 

This range covers a large variety of  foams already studied in the literature. Larger $\Delta n/ \langle n  \rangle $, up to 0.5 or 0.6, would instead correspond to simulated  irregular-shaped  structures  generated by fragmentation  (fractal topological gas  \cite{delannay}). 
Lower $\Delta n/ \langle n  \rangle $, down to zero, can be obtained only in highly controlled structures, with
periodic boundary conditions (as in simulations) or with 
 boundaries far away from the analysed image
  \cite{Theory}.
  For instance  the fruit fly retina (Fig  \ref{fig:complete}b) is part of a  large honeycomb lattice of 800 facets
  \cite{retina}.
  
According to de Almeida and coworkers 
\cite{rita2D,rita3D,thomas}, ``shuffling" a foam involves enough
topological changes (including ``T1" neighbour swappings \cite{weairerivier})  
that it loses any memory of its initial condition. 
The present results make this definition precise: a foam is shuffled after enough rearrangements   \cite{rau07} that the topological disorder becomes more dependent on the geometrical disorder than on the initial conditions.
Since the present work is descriptive, it is difficult to determine the physical origin 
of this  statistical (rather than exact) correlation, which is not trivial  \cite{Theory}.
It does not seem related to the minimisation of any energy, and could apply to patterns other than foams. 

Examples 
 from foam simulations
 as well as biological tissues 
 are plotted on Fig. \ref{fig:complete}. The collected data all fall in the same region of the (geometrical disorder, topological disorder) plane.
Most of them have
  $\Delta n/ \langle n \rangle$
   larger (up to 1.5 or 2 times) 
  than eq. (\ref{fit_ecart_type}).

The curve   of Fig. \ref{cath_dav} (eq. \ref{fit_ecart_type})   appears, again  statistically 
(rather than exactly \cite{Theory}),
 as a lower bound for the topological disorder, at given area disorder.  It can be used in practice as a benchmark to test how shuffled a pattern is.
If a pattern has  $\Delta n/ \langle n \rangle$ very different from $0.3 \Delta A/ \langle A \rangle$,
 it probably means that it is far from being shuffled. There is no reciprocity: a non-shuffled pattern can be close to eq. (\ref{fit_ecart_type}). 
The curve   of Fig. \ref{cath_dav} 
seems to be stable under shuffling:  it is reached asymptotically, when starting from different initial conditions (Fig. \ref{Deltan_vs_t}).

\section{Conclusion}

To summarize, all the patterns that we examine are in the same region of the plane 
(geometrical disorder, topological disorder).
For a given width of the area distribution (and independently of its higher moments), the ``shuffled foam" is well defined and realized. It reaches a lower bound  (eq. \ref{fit_ecart_type}) for the topological disorder, which is robust and directly dependent on the geometrical disorder.
Theoretical work could try to explain eq. (\ref{fit_ecart_type}) and its stability, for instance by understanding the respective roles of  disorder and surface minimisation \cite{rita2D,rita3D,thomas}. 

These results could be used to characterize patterns and determine causal relations between both disorders, including in biological mutants.  They could also be used to study how the foam's disorder (now quantified by a single number) affects its mechanical properties
\cite{simon_elong}, and to improve the characterisation of coarsening \cite{glazierweaire,thomas}.
Perspectives include generalisation to other discrete systems, such as granular matter and colloids \cite{ziherl,likos}
and to 3D systems \cite{thomas}.
    
\section*{Acknowledgements}

Manuel Fortes' passionate approach to foam physics has been a source of inspiration for several of us. We thank J. Legoupil for his participation in the simulations, K. Brakke for developing and maintaining the Surface Evolver code, A.F.M. Mar\'ee for developing the Potts model code used here, M.F. Vaz for providing references, I. Cantat for critical reading of the manuscript,
Y. Bella\"{i}che, R. Carthew and T. Hayashi for providing pictures of biological tissues, S. Courty for their analysis, P. Ballet for help in setting up the experiment, and participants of the Foam Mechanics workshop (Grenoble, January 2008) for many discussions.
S.A.T. thanks Dr Ejtehadi for hospitality at the Institute of Physics and Mathematics, Tehran (Iran). S.J.C. thanks UJF for hospitality, and CNRS, EPSRC (EP/D048397/1, EP/D071127/1) and the British Council Alliance scheme for financial support.


\begin{thebibliography}{1}
\bibitem{Theory}
For any $ \Delta A/ \langle A \rangle$ it would be possible to construct a (non-shuffled!) foam of arbitrarily small  $\Delta n/ \langle n  \rangle$, as follows  \protect\cite{teixeira}.
Consider
 a  bimodal foam, where small bubbles, arranged in a honeycomb, would be surrounded by a honeycomb of large bubbles. 
Non-hexagonal bubble would be at the boundary between both honeycombs, and thus of order $\sqrt{N}$. Thus  by increasing 
$N$,  $\Delta n/ \langle n  \rangle$ could be made arbitrarily small.
\end{thebibliography}
\end{document}